\documentclass[runningheads]{llncs}
\usepackage[T1]{fontenc}
\usepackage{graphicx}
\usepackage{comment}

\usepackage{xcolor}
\usepackage{hyperref}
\usepackage{booktabs}
\usepackage{multirow}
\usepackage{amssymb}
\usepackage{soul}
\usepackage{multirow}
\usepackage{tabularx}

\newcommand\JD[1]{\textcolor{black}{#1}}

\usepackage{fontawesome5}
\usepackage{mdframed}

\mdfdefinestyle{stebox2}{
	 backgroundcolor=gray!10,
	linecolor=gray!50,
	roundcorner=5pt,
	leftmargin=0pt,
	rightmargin=0pt,
	skipabove=3pt,
	skipbelow=3pt
}

\newcommand{\rqanswerfirst}[1]{%
	\begin{mdframed}[style=stebox2]
		\textbf{{Answer to RQ1:}} #1
	\end{mdframed}
}

\newcommand{\rqanswersec}[1]{%
	\begin{mdframed}[style=stebox2]
		\textbf{{Answer to RQ2:}} #1
	\end{mdframed}
}

\newcommand{\rqanswerthree}[1]{%
	\begin{mdframed}[style=stebox2]
		\textbf{{Answer to RQ3:}} #1
	\end{mdframed}
}

\begin{document}

\title{From Online User Feedback to Requirements: Evaluating Large Language Models for Classification and Specification Tasks}
\titlerunning{From Online User Feedback to Requirements: An Empirical Study}
%

\author{Manjeshwar Aniruddh Mallya\inst{1} \and
Alessio Ferrari\inst{2}\orcidID{0000-0002-0636-5663} \and
Mohammad Amin Zadenoori\inst{3}\orcidID{0000-0003-4591-153X}
\and \\
Jacek Dąbrowski\inst{1}\orcidID{0000-0003-3392-0690}
}
\authorrunning{A. M. Mallya et al.}
%

\institute{
	Lero, the Research Ireland Centre for Software, University of Limerick, Ireland \\
	\email{\{24124133@studentmail.ul.ie,jacek.dabrowski\}@lero.ie} \and
	University College Dublin (UCD), Ireland \\
	\email{alessio.ferrari@ucd.ie} \and
	University of Padova, Italy \\
	\email{amin.zadenoori@unipd.it}
}

\maketitle              

\begin{abstract} 
	
	\textbf{[Context and Motivation]} Online user feedback provides valuable information to support requirements engineering (RE). However, analysing online user feedback is challenging due to its large volume and noise. Large language models (LLMs) show strong potential to automate this process and outperform previous techniques. They can also enable new tasks, such as generating requirements specifications. \textbf{[Question/Problem]} Despite their potential, the use of LLMs to analyse user feedback for RE remains underexplored. Existing studies offer limited empirical evidence, lack thorough evaluation, and rarely provide replication packages, undermining validity and reproducibility. \textbf{[Principal Idea/Results]} \JD{We evaluate five lightweight open-source LLMs on three RE tasks: user request classification, NFR classification, and requirements specification generation. Classification performance was measured on two feedback datasets, and specification quality via human evaluation. \JD{LLMs achieved moderate-to-high classification accuracy (F1 $\approx$ 0.47–0.68) and moderately high specification quality (mean $\approx$ 3/5).}} \textbf{[Contributions]} \JD{We newly explore lightweight LLMs for feedback-driven requirements development.} Our contributions are: (i) an empirical evaluation of lightweight LLMs on three RE tasks, (ii) a replication package, and (iii) insights into their capabilities and limitations for RE.
	
	\keywords{Requirements Engineering \and User Feedback  \and Empirical Study \and AI4RE \and NLP \and Mining Software Repository \and Large Language Models.}
\end{abstract}

\section{Introduction}
\label{sec:introduction}

Online user feedback is a valuable source of information that supports various requirements engineering (RE) tasks~\cite{Dabrowski2022a,Ferrari2022}. Such feedback often provides insights into bug reports, feature requests, and non-functional requirements (NFRs)~\cite{Dabrowski2022}. However, manually analyzing this feedback is difficult because of its large volume and noisy nature~\cite{AlSubaihin2021,11190331}. Automating this analysis could help practitioners identify user needs and document requirements more effectively.

Numerous approaches have been proposed to analyze user feedback using natural language processing (NLP) and machine learning (ML) techniques for tasks such as classification and opinion mining~\cite{Dabrowski2022,Dabrowski2023}. However, these methods often overlook key RE tasks, such as generating requirements specifications; \JD{their effectiveness remains limited in tasks like NFR categorization, essential for addressing software quality attributes such as security and usability~\cite{4384163}}.

Recent advances in large language models (LLMs) create new opportunities to address these limitations~\cite{Arora2024,Fan2023,11190385}. LLMs perform well in classification~\cite{alhoshan2025effectivegenerativelargelanguage}, summarization, and text generation~\cite{hou2024large,cheng2024generative}. Their use in RE has been explored for tasks such as traceability and ambiguity detection but remains limited \JD{for leveraging user feedback, particularly for generating requirements specifications}~\cite{zadenoori2025largelanguagemodelsllms}. Most existing studies rely on commercial, resource-intensive models and lack systematic evaluation or replication packages, reducing reproducibility~\cite{Abualhaija2024,Dabrowski2022}.

The goal of this research is to address prior limitations and explore lightweight open-source LLMs for automating online user feedback analysis in support of RE tasks. \JD{More specifically, we empirically evaluate five lightweight open-source LLMs using five prompting strategies on three RE tasks:} classifying feedback by user request type, classifying by NFR type, and generating requirement specifications. Classification uses two annotated user feedback datasets, while specification quality is assessed through human evaluation.

The main contributions are: (i) an empirical study of lightweight LLMs on three RE tasks using user feedback, (ii) a replication package~\cite{mallya2026from}, and (iii) insights into the capabilities and limits of lightweight LLMs in RE. \JD{To our knowledge, this is the first empirical study of lightweight LLMs for feedback-driven requirements development, supported by a replication package~\cite{Dabrowski2022,zadenoori2025largelanguagemodelsllms}.}

The remainder of this paper is as follows. Sect.~\ref{sec:background} introduces the terminology and problem, followed by an overview of the selected LLMs and prompting strategies. Sect.~\ref{sec:motivation} presents the motivating scenarios, Sect.~\ref{sec:design} details the study design, and Sect.~\ref{sec:results} reports the results. Sect.~\ref{sec:discussion} discusses the findings and, Sect.~\ref{sec:Threatstovalidity} details threats to validity. Sect.~\ref{sec:RelatedWorks} reviews related work, and Sect.~\ref{sec:conculsion} concludes the paper.
\section{Background}
\label{sec:background}

We now define key terms, formulate the problems of user feedback classification and requirements specification; and outline approaches we selected in our study.

\subsection{Terminology and Problem Formulation}

\textbf{Definition 1 (User Feedback).} A user feedback instance is denoted as \( r \), representing a textual message written by a user about an application. 
The set of all feedback messages for an application \( a \) is \( R = \{r_1, r_2, ..., r_n\} \).

This study focuses on app reviews, a form of user feedback from mobile platforms (e.g., Google Play Store). Such feedback conveys diverse information useful for RE; in this work, we particularly focus on user requests and  NFRs.

\noindent\textbf{Definition 2 (User Request).} 
A user request \( t_{UR} \) is an expressed intent in a feedback instance, which can be of a type \textit{feature request}, or \textit{bug report}.

\noindent\textbf{Definition 3 (Non-Functional Requirement).} 
A non-functional requirement \( t_{NFR} \) is a reference in a feedback instance to one of the eight quality attributes of the application (e.g., \textit{usability}) as defined in ISO/IEC 25010~\cite{iso25010}.

Our study addresses two main problems: user feedback classification and requirements specification, as defined below. 

\noindent\textbf{Problem 1 (User Feedback Classification).} 
Given a set of user feedback instances \( R = \{r\} \) for an application \( a \), find a multi-set \( C = \{t\} \), where \( t \) is an information type assigned to a feedback instance \( r \in R \). The type classification is performed either on user-request types \( t_{UR} \) or on NFR types \( t_{NFR} \), each with an additional \textit{other} category to capture feedback outside these types.

\noindent\textbf{Problem 2 (Requirements Specification Generation).} 
Given a set of user feedback instances \( R = \{r\} \) for an application \( a \) and their associated information types \( C = \{t\} \), generate a set of requirement statement \( S = \{s\} \), where each \( s \) is a written statement that formalizes or summarizes the user needs expressed in one or more feedback instances \( r \in R \). 

\subsection{Large Language Models}

\textit{Large Language Models (LLMs)} are neural networks trained on large text corpora~\cite{Hou2024,Fan2023}. They show strong performance in reasoning, summarization, and text generation~\cite{Arora2024,zadenoori2025largelanguagemodelsllms}. Commercial models such as GPT-4, Claude, and Gemini are widely used in both industry and research. However, they are resource-intensive, provider-dependent, and difficult to customize. These characteristics limit their suitability for controlled research and small-scale SE projects.

\textit{Lightweight open-source LLMs} provide a practical alternative. They contain fewer parameters and require less computational power. As a result, they can run efficiently on local machines. Their open and adaptable nature also makes them suitable for small-company projects and open-science research. Recent advances in such models have created new opportunities to explore their potential in RE.

In this study, we focus on five open-source lightweight LLMs: Llama 2, Llama 3, Mistral, Gemma 2, and Phi-3 Mini. For simplicity, we refer to these lightweight LLMs simply as LLMs throughout the paper. \JD{We selected these models from major AI developers as a representative set of open-source lightweight LLMs for RE experimentation. They differ in size and reasoning capability but can all be executed locally. The aim was to examine their performance and behavior rather than identify the best model.} Table~\ref{tab:llms} summarises the LLMs used in our study. Model refers to the model name, and Developer to the releasing organisation. Parameters (in billions) indicate the number of trainable weights, reflecting each model’s capacity and computational cost. The Context window shows how much text the model can process in one input, measured in tokens; larger windows support longer and more coherent inputs. Usage distinguishes models available for open access from those restricted to research.

\begin{table}[t]
	\centering
	\caption{Lightweight open-source LLMs used in our study.}
	\setlength{\tabcolsep}{4.5pt} 
	\scalebox{0.8}{
		\begin{tabular}{|l|l|c|c|l|}
			\hline
			\textbf{Model} & \textbf{Developer} & \textbf{Parameters} & \textbf{Context Window} & \textbf{Usage} \\
			\hline
			Llama 2 & Meta AI & 7B & 4K & Research \\
			Llama 3 & Meta AI & 8B & 8K & Research \\
			Mistral & Mistral AI & 7B & 8K & Open \\
			Gemma 2 & Google DeepMind & 9B & 8K & Open \\
			Phi-3 Mini & Microsoft & 3.8B & 4K & Open \\
			\hline
		\end{tabular}
	}
	\label{tab:llms}
\end{table}

Interaction with LLMs occurs through \textit{prompts}~\cite{binkhonain2025prompts}, which are textual instructions that define the task and expected output~\cite{zadenoori2025largelanguagemodelsllms}. The way a prompt is formulated, referred to as a \textit{prompt strategy}, strongly influences the quality of the generated results~\cite{vogelsang2024usinglargelanguagemodels}. Our study examines five representative strategies, as shown in Table~\ref{tab:prompt_strategies}. The Zero shot approach uses only task descriptions, while Few shot prompting provides a small set of labeled examples. Chain of thought prompts encourage step-by-step reasoning, which improves interpretability. Constraint based prompts introduce explicit rules to ensure structured and consistent outputs. Finally, Role based prompts assign the model a specific persona so that its responses align with relevant professional or contextual expectations.

\begin{table}[h!]
	\centering
	\caption{Overview of prompt strategies used in our study.}
	\label{tab:prompt_strategies}
	\scalebox{0.8}{%
		\begin{tabular}{|p{3cm}|p{10.5cm}|}
			\hline
			\textbf{Prompt Strategy} & \textbf{Description and Example} \\
			\hline
			\multirow{4}{*}{\centering\textbf{Zero-shot}} &
			\textbf{Description:} A brief instruction describes the task, assuming the model can generalise from its pre-trained knowledge to perform it. \\
			\cline{2-2}
			& \textbf{Example:} ``Classify the following user feedback as a feature request, bug report, or usability issue.'' \\
			\hline		
			\multirow{4}{*}{\centering\textbf{Few-shot}} &
			\textbf{Description:} A few labelled examples show the desired pattern, assuming the model will apply it to new inputs. \\
			\cline{2-2}
			& \textbf{Example:} ``Example 1: `Add dark mode' $\rightarrow$ Feature Request. 
			Example 2: `App crashes on login' $\rightarrow$ Bug Report. 
			Now classify the next five items.'' \\
			\hline
			\multirow{4}{*}{\centering\textbf{Chain-of-thought}} &
			\textbf{Description:} The model is instructed to reason step by step and write intermediate steps before giving the final answer. \\
			\cline{2-2}
			& \textbf{Example:} ``Explain why this user comment indicates a performance issue, then label it as an NFR.'' \\
			\hline
			\multirow{4}{*}{\centering\textbf{Constraint-based}} &
			\textbf{Description:} Prompts include explicit rules, templates, or conditions that the output must follow. \\
			\cline{2-2}
			& \textbf{Example:} ``Generate requirements that are testable, unambiguous, and measurable.'' \\
			\hline
			\multirow{4}{*}{\centering\textbf{Role-based}} &
			\textbf{Description:} The model is assigned a specific role and responds using the perspective, tone, and knowledge expected from it. \\
			\cline{2-2}
			& \textbf{Example:} ``You are a requirements analyst. Rewrite the following user request as a formal functional requirement.'' \\
			\hline
		\end{tabular}
	}
\end{table}
\section{Motivating Scenarios}
\label{sec:motivation}

We present three scenarios illustrating how feedback classification and requirements generation support RE, inspired by real cases~\cite{AlSubaihin2021}, and prior studies~\cite{Dabrowski2022}.

\noindent\textbf{Scenario 1 (Supporting Requirements Elicitation).} App reviews often contain implicit requirements expressed as feature requests or bug reports~\cite{Dabrowski2022a}. Understanding these helps product teams capture evolving user needs, identify emerging issues, and plan product improvements. As an example, imagine that after releasing a new version of WhatsApp, the team wants to identify problems users mention in their reviews (e.g., ``messages fail to send'') and desired features (e.g., ``add message scheduling''). Automatically classifying app reviews by request type helps teams quickly identify user needs, while quantifying the feedback shows their user-perceived importance.

\noindent\textbf{Scenario 2 (Supporting Requirements Classification).} Beyond identifying request types, the WhatsApp development team also wants to understand which non-functional qualities users discuss~\cite{Dabrowski2022}. Feedback such as ``too slow to open'' or ``confusing chat layout'' reflects concerns about performance and usability. Automatically classifying feedback by referenced NFRs helps developers identify which quality attributes users value most, revealing whether users are more concerned with performance or usability. This supports task prioritization and helps balance new features with quality improvements.

\noindent\textbf{Scenario 3 (Supporting Requirements Specification).} After identifying and classifying requirements from WhatsApp user feedback, analysts may want to document them clearly and consistently. Translating informal comments into structured requirements is time-consuming and error-prone~\cite{AlSubaihin2021}. For example, feedback such as ``add dark mode'' can be reformulated as ``The system shall provide a dark mode option''. Although such documentation may not produce complete specifications, automating it can provide initial drafts that accelerate refinement and maintain traceability with user feedback.

For these scenarios, a tool that classifies user feedback and generates requirements could help the team evolve their app more effectively.

\section{Empirical Study Design}
\label{sec:design}

This section presents the empirical study conducted to evaluate the effectiveness of LLMs in analysing online user feedback to support RE. 

\subsection{Research Questions}
\label{sec:RQs}
The goal of this study is to evaluate LLMs in analysing online user feedback to support RE tasks. We specifically focus on three research questions:

\begin{itemize}
	\item \textbf{RQ1}: How well do LLMs classify feedback by NFRs type?
	\item \textbf{RQ2}: How well do LLMs classify feedback by user-request type?
	\item \textbf{RQ3}: How well do LLMs generate requirements specifications?
\end{itemize}

In RQ1, we assess the models’ ability to identify the NFR type mentioned in user reviews. RQ2 examines how accurately the models classify user feedback by request type. RQ3 evaluates their capability to generate requirements specifications from the same feedback. All evaluations use human-annotated datasets (see Sect.~\ref{sec:datasets}). For RQ1–RQ2, model predictions are compared with annotations using precision, recall, and F1-score. For RQ3, SRSs generated from sampled reviews are assessed through human judgment based on predefined quality criteria.

\subsection{Datasets}
\label{sec:datasets}

We use two annotated datasets of mobile app user feedback from prior studies~\cite{lu2017automatic}. We selected them for their relevance and inclusion in a public repository~\cite{Dabrowski2022}. Each contains thousands of reviews from about a dozen apps across varied domains and both major app stores. This diversity mitigates the app sampling problem and supports validity~\cite{Dabrowski2022}.

\noindent\textbf{User Request Dataset.} The first dataset builds on previous studies~\cite{chen2014arminer,maalej2016on} and includes additional user feedback collected for this study~\cite{jha2018using}. The initial collected dataset covered about 10 mobile applications across more than 10 categories from both the Apple App Store and Google Play Store. From an initial pool of the collected reviews, a curated subset of 2,912 was manually annotated into three user request types: feature requests, bug reports, and other. This dataset provides a diverse and representative sample of user feedback for evaluating automated classification approaches

\noindent\textbf{Non-Functional Requirements (NFR) Dataset.} The second dataset, introduced by Lu and Liang~\cite{lu2017automatic}, focuses on app reviews annotated with NFR types based on the software quality model~\cite{iso25010}. The initial collected dataset comprised approximately 11,000 app reviews from two mobile applications in the books and communications categories. These reviews were drawn from both major app stores, the Apple App Store and Google Play Store. A subset of 4,000 review sentences was manually annotated according to five NFR categories.

\subsection{Evaluation Metrics and Criteria}
\label{sec:metrics}

We applied both quantitative and qualitative methods. Standard ML metrics~\cite{11190353} were used for RQ1 and RQ2, as feedback classification is a classification task~\cite{geron2022hands}. Specification generation (RQ3) was evaluated via criteria-based assessment~\cite{kuckartz2014qualitative}.

\paragraph{Evaluation Metrics for Classification (RQ1--RQ2).} We compute precision, recall, and F1-score for two experiments: classifying user feedback by NFR type (RQ1) and by user request type (RQ2). Precision measures the proportion of correctly predicted labels among all predictions, while recall measures the proportion of correctly identified labels in the ground truth. Model performance is assessed by comparing each predicted label (user request or NFR type) with its annotated counterpart. Scores are calculated per class (e.g., feature request, bug report, other), along with macro averages. The macro average treats all classes equally.

\paragraph{Evaluation Criteria for Specification Generation (RQ3).} For RQ3, the quality of generated requirement specifications is evaluated qualitatively across six criteria derived from established specification attributes such as completeness, consistency, and clarity~\cite{10834143}. Each criterion was adapted to suit the characteristics of automatically generated specifications. Table~\ref{tab:srs_rubric} outlines the evaluation rubric, which includes six criteria. These dimensions assess structural correctness, coverage of stakeholder input, factual grounding, semantic accuracy, and linguistic quality. Each criterion follows a defined scoring scheme combining quantitative (e.g., counts or coverage) and qualitative (e.g., 1–5 scale) measures, supporting systematic and replicable assessment of specification quality~\cite{kuckartz2014qualitative}.

\begin{table}[t]
	\centering
	\caption{\JD{Evaluation criteria and scoring used to assess generated specification (RQ3).}}
	\label{tab:srs_rubric}
	\scalebox{0.8}{%
		\begin{tabular}{|p{3.6cm}|p{10cm}|}
			\hline
			\textbf{Criterion} & \textbf{Description and Scoring} \\
			\hline
			
			\multirow{4}{*}{\centering\textbf{Structural Adherence}} &
			\textbf{Description:} Evaluates how well the generated specification follows the expected structure, ensuring coverage of all sections. \\
			\cline{2-2}
			& \textbf{Scoring:} \JD{Originally 1–8 points (one per correctly included section), linearly rescaled to 1–5 for consistency; higher is better.} \\
			\hline
			
			\multirow{4}{*}{\centering\textbf{Completeness}} &
			\textbf{Description:} Evaluates the coverage of requirements identified from user feedback. \\
			\cline{2-2}
			& \textbf{Scoring:} Rated on a 1–5 scale, with higher scores reflecting greater completeness of identified requirements from user feedback. \\
			\hline
			
			\multirow{4}{*}{\centering\textbf{Fidelity}} &
			\textbf{Description:} \JD{Evaluates how faithfully the generated specification reflects user feedback, identifying fabricated requirements.} \\
			\cline{2-2}
			&  \JD{\textbf{Scoring:} Rated on a 1–5 scale; higher scores indicate greater fidelity (rescaled from the proportion of fabricated to valid requirements).}	\\
			\cline{2-2}
			\hline
			\multirow{4}{*}{\centering\textbf{Conciseness}} &
			\textbf{Description:} Evaluates redundancy and verbosity in the generated text, indicating how efficiently information is conveyed. \\
			\cline{2-2}
			& \textbf{Scoring:} \JD{Rated on a 1–5 scale, with higher scores reflecting greater conciseness and lower values indicate increased verbosity.}\\
			\hline
			\multirow{4}{*}{\centering\textbf{Clarity}} &
			\textbf{Description:} Evaluates the clarity of generated requirements in terms of unambiguity, specificity, and interpretability. \\
			\cline{2-2}
			& \textbf{Scoring:} Rated on a 1–5 scale, with higher scores reflecting greater clarity and linguistic precision. \\
			\hline
			
		\end{tabular}
	}
\end{table}

\subsection{Experimental Setup and Procedure}
\label{sec:setup}

We now describe the computational setup, prompting strategies, and evaluation procedures used in three experiments (RQ1--RQ3).

\subsubsection{Computational Setup}
\label{sec:comp-setup}
All experiments were conducted under consistent hardware and parameter settings to ensure fairness and reproducibility. We evaluated five LLMs (see Sect.~\ref{sec:background}), running each model three times per task to reduce stochastic variance. Default parameters were used, with \textit{temperature} set to 0 and a fixed \textit{random seed} to minimise non-determinism. No hyperparameter tuning was applied to isolate the effects of prompting strategies. Experiments ran on a workstation with an NVIDIA RTX 4050 GPU (6 GB VRAM) and 16 GB RAM. Total runtime per model (3 runs) ranged from 20 min to 2 h for user request classification (512 reviews) and 2–5 hours for NFR classification (1,278 reviews).

\subsubsection{Prompting Strategies}
\label{sec:prompts} 

We experiment with five prompting strategies (see Sect.~\ref{sec:background}). Prompting strategies are customised for each experiment (RQ1--RQ2).

\paragraph{Prompting Strategies for  Classification (RQ1--RQ2).} We apply three prompting strategies across two classification tasks (user request type and NFR type): \textit{zero-shot}, \textit{few-shot}, and \textit{chain-of-thought (CoT)} prompting. These strategies capture increasing levels of reasoning and contextualization while remaining lightweight and reproducible. We omit more complex prompting (e.g., role-based or constraint-based), as classification primarily requires consistent label prediction rather than creative or constrained generation. Prompts are refined iteratively through a pilot study. Few-shot examples come from our dataset, and CoT prompts direct models to ``\textit{think step-by-step before giving the final category}''.

\paragraph{Prompting Strategies for  Specification Generation (RQ3).} We use all five prompting strategies for the specification generation task. Beyond the three classification strategies (\textit{zero-shot}, \textit{few-shot}, \textit{CoT}), we add \textit{constraint-based} and \textit{role-based} prompting to improve structural coherence and contextual relevance. These strategies better suit generative tasks that require creativity and controlled output. In the constraint-based setup, prompts specify that outputs follow a defined structure (e.g., functional and non-functional sections) and include constraints such as ``\textit{avoid implementation details}'' and ``\textit{ensure each requirement is unique}''.

\subsubsection{Evaluation Procedures}
\label{sec:eval-proc}

We use quantitative evaluation for classification tasks (RQ1--RQ2) and qualitative evaluation for specification generation (RQ3).

\paragraph{Evaluation Procedure Classification (RQ1--RQ2).} For RQ1 and RQ2, we evaluate models on the corresponding annotated dataset for each classification experiment (Section~\ref{sec:datasets}). Each review is processed by the LLM under each prompting setup (zero-shot, few-shot, CoT), and predicted labels are compared with the ground truth. We calculate precision, recall, and F1-score per class, along with macro- and weighted averages. We report mean values over three runs.

\paragraph{Evaluation Procedure for Specification Generation (RQ3).} For RQ3, we use 90 annotated reviews sampled from our collected data. Half are taken from the user request dataset, and the other half from the NFR dataset. The same input is provided to each LLM under every prompting setup. Each model generates requirements specifications following the template, including Introduction, Functional Requirements, NFRs, and Glossary sections. \JD{The first author evaluates the outputs using the five criteria (see Table~\ref{tab:srs_rubric}). The results are verified through manual examination of user feedback. The scores are averaged across all samples.}

\section{Results}
\label{sec:results}

\noindent\textbf{RQ1: How well do LLMs classify feedback by NFR type?}

\noindent To answer RQ1, we evaluated how well LLMs classify user feedback by NFRs under three prompting strategies: zero-shot, few-shot, and chain-of-thought. Table \ref{tab:llm-nfrs-performance} reports precision, recall, and F1 scores for each model, with the best results highlighted in bold. The effectiveness ranges from an F1 score of 0.40 to 0.55, with averages of 0.47, 0.49, and 0.51 for the zero-shot, few-shot, and chain-of-thought strategies, respectively.  Across prompting strategies, performance improves steadily from zero-shot to few-shot to chain-of-thought, confirming that example-based and reasoning-enhanced prompts help LLMs better identify NFR types. Larger and newer models (Gemma, Llama 3, Mistral) generally outperform smaller or earlier ones (Llama 2, Phi-3 Mini). Gemma achieves the highest overall F1 score (0.55) and precision (0.53) under the chain-of-thought setting, while Llama 3 records the highest recall (0.59). Mistral performs most consistently across all prompting strategies, ranking near the top in zero-shot and few-shot modes. In contrast, Llama 2 yields the lowest scores across all metrics, while Phi-3 Mini remains stable but below the larger models.

\rqanswerfirst{LLMs achieve moderate accuracy (F1 $\approx$  0.47–0.51) for classifying feedback by NFR type. Chain-of-thought prompting performs best, with Gemma leading overall (P=0.53, R=0.57, F1=0.55).}

\begin{table*}[h]
	\centering
	\caption{How well do LLMs classify feedback by NFR type? (RQ1)}
	\label{tab:llm-nfrs-performance}
	\setlength{\tabcolsep}{6pt} 
	\renewcommand{\arraystretch}{1.2} 
	\scalebox{0.8}{
		\begin{tabular}{|l|c|c|c|c|c|c|c|c|c|}
			\hline
			\multirow{2}{*}{\textbf{Model}} 
			& \multicolumn{3}{c|}{\textbf{Zero-Shot}} 
			& \multicolumn{3}{c|}{\textbf{Few-Shot}} 
			& \multicolumn{3}{c|}{\textbf{Chain-of-Thought}} \\
			\cline{2-10}
			& \textbf{P} & \textbf{R} & \textbf{F1}
			& \textbf{P} & \textbf{R} & \textbf{F1}
			& \textbf{P} & \textbf{R} & \textbf{F1} \\
			\hline
			Llama 2        & 0.44 & 0.36 & 0.40 & 0.49 & 0.39 & 0.43 & 0.52 & 0.46 & 0.49 \\
			Llama 3        & 0.42 & \textbf{0.54} & 0.47 & 0.46 & \textbf{0.57} & 0.51 & 0.48 & \textbf{0.59} & 0.53 \\
			Mistral        & 0.49 & 0.51 & \textbf{0.50} & \textbf{0.54} & 0.51 & \textbf{0.52} & 0.47 & \textbf{0.59} & 0.52 \\
			Gemma          & \textbf{0.51} & 0.48 & 0.49 & 0.51 & 0.49 & 0.50 & \textbf{0.53} & 0.57 & \textbf{0.55} \\
			Phi-3 Mini     & 0.44 & 0.54 & 0.48 & 0.46 & 0.54 & 0.50 & 0.44 & 0.48 & 0.46 \\
			\hline
			\textbf{Average} & 0.46 & 0.49 & 0.47 
			& 0.49 & 0.50 & 0.49 
			& 0.49 & 0.54 & 0.51 \\
			\hline
		\end{tabular}
	}
\end{table*}

\noindent \textbf{RQ2: How well do LLMs classify feedback by user request type?}

\noindent To answer RQ2, we examined how LLMs classify user feedback by request type under three prompting strategies: zero-shot, few-shot, and chain-of-thought. Table \ref{tab:llm-ur-performance} presents precision, recall, and F1 scores for each model, with the best results in bold. Model performance ranges from an F1 score of 0.32 to 0.74. The average F1 values are 0.59 for zero-shot, 0.68 for few-shot, and 0.64 for chain-of-thought prompting. Performance increases notably from zero-shot to few-shot prompting, while chain-of-thought yields moderate improvements. Larger and more recent models, such as Llama 3, Mistral, and Gemma, generally achieve higher accuracy than smaller or earlier ones, including Llama 2 and Phi-3 Mini. Llama 3 reaches the highest F1 score of 0.74 and recall of 0.75 under the few-shot setting. Mistral and Gemma perform consistently well across all strategies. Llama 2 produces the lowest scores, while Phi-3 Mini remains stable but below the stronger models. Overall, few-shot prompting provides the best results (average F1 = 0.68); it suggests that including example-based context helps LLMs more accurately classify user feedback by request type.

\rqanswersec{LLMs achieve moderate-to-high average accuracy (F1 $\approx$ 0.59–0.68) for classifying feedback by user request type. Few-shot prompting performs best, with Llama 3 leading overall (P=0.72, R=0.75, F1=0.74).}

\begin{table*}[h]
	\centering
	\caption{How well do LLMs classify user feedback by request type? (RQ2)}
	\label{tab:llm-ur-performance}
	\setlength{\tabcolsep}{6pt} 
	\renewcommand{\arraystretch}{1.2} 
	\scalebox{0.8}{
		\begin{tabular}{|l|c|c|c|c|c|c|c|c|c|}
			\hline
			\multirow{2}{*}{\textbf{Model}} 
			& \multicolumn{3}{c|}{\textbf{Zero-Shot}} 
			& \multicolumn{3}{c|}{\textbf{Few-Shot}} 
			& \multicolumn{3}{c|}{\textbf{Chain-of-Thought}} \\
			\cline{2-10}
			& \textbf{P} & \textbf{R} & \textbf{F1}
			& \textbf{P} & \textbf{R} & \textbf{F1}
			& \textbf{P} & \textbf{R} & \textbf{F1} \\
			\hline
			Llama 2        & 0.28 & 0.36 & 0.32 & 0.57 & 0.56 & 0.57 & 0.60 & 0.42 & 0.49 \\
			Llama 3        & \textbf{0.77} & 0.67 & \textbf{0.72} & 0.72 & 0.75 & \textbf{0.74} & 0.71 & 0.70 & 0.71 \\
			Mistral        & 0.60 & 0.63 & 0.62 & 0.69 & \textbf{0.74} & 0.71 & 0.65 & \textbf{0.72} & 0.68 \\
			Gemma          & 0.66 & \textbf{0.71} & 0.68 & 0.68 & 0.67 & 0.68 & 0.65 & 0.60 & 0.63 \\
			Phi-3 Mini     & 0.69 & 0.51 & 0.59 & 0.68 & 0.67 & 0.68 & 0.67 & 0.70 & \textbf{0.69} \\
			\hline
			\textbf{Average} & 0.60 & 0.58 & 0.59 
			& 0.67 & 0.68 & 0.68 
			& 0.66 & 0.63 & 0.64 \\
			\hline
		\end{tabular}
	}
\end{table*}

\noindent \textbf{RQ3: How well do LLMs generate requirements specifications?}

\noindent To answer RQ3, each model was evaluated on five criteria: structure (SA), completeness (CO), fidelity (FI), conciseness (CN), and clarity (CL), as defined in Table~\ref{tab:srs_rubric}. Table~\ref{tab:llm-srs-performance} reports results across models and prompting strategies. Overall, the models produced moderate-quality specifications (mean=3.1; SD=0.8). Llama 3 with chain-of-thought prompting and Mistral with few-shot prompting achieved the highest mean score (3.6), showing strong structure and clarity (SA=4; CL=5). Llama 2 followed with stable mid-range scores (mean=3.2–3.4). Gemma produced the weakest but most consistent outputs (mean $\approx$ 2.9; SD=0.4). Phi-3 Mini showed high fidelity and conciseness (FI $\approx$ 4; CN=4–5) but low structure (SA=1–2) and high variability. Across prompting strategies, few-shot and chain-of-thought improved structure and clarity, whereas constrained-based prompting offered limited gains. Most models scored highest in clarity (CL=4–5) and completeness (CO=4) but lagged with fidelity and conciseness (FI, CN=2–3).

\rqanswerthree{LLMs produced moderate-quality specifications, with Llama 3 and Mistral performing best; model and prompt choice strongly influenced output quality.}

\begin{table*}[h]
	\centering
\caption{LLM performance on requirement specifications (RQ3). 
	SA – Structure; CO – Compl.; FI – Fidelity; CN – Conciseness; CL – Clarity. 
	Scores are on a 1–5 scale (higher = better). 
	\textbf{Bold} marks the highest values across all models and prompt strategies.}

	\label{tab:llm-srs-performance}
	\setlength{\tabcolsep}{4pt}
	\renewcommand{\arraystretch}{1.2}
	\scalebox{0.76}{
		\begin{tabular}{|l|l|c|c|c|c|c|c|c|}
			\hline
			\textbf{Model} & \textbf{Prompt Strategy} & 
			\textbf{SA (1–5)} & 
			\textbf{CO (1–5)} & 
			\textbf{FI (1–5)} & 
			\textbf{CN (1–5)} & 
			\textbf{CL (1–5)} & 
			\textbf{Mean} & 
			\textbf{SD} \\ 
			\hline
			
			\multirow{5}{*}{Llama 2} 
			& Zero-Shot & 3 & 4 & 3 & 3 & 4 & 3.4 & 0.49 \\ \cline{2-9}
			& Few-Shot & 3 & 4 & 3 & 2 & 4 & 3.2 & 0.75 \\ \cline{2-9}
			& Chain-of-Thought & 3 & 4 & 3 & 2 & 4 & 3.2 & 0.75 \\ \cline{2-9}
			& Constraint-based & 3 & 4 & 3 & 2 & 4 & 3.2 & 0.75 \\ \cline{2-9}
			& Role-based & 3 & 4 & 3 & 2 & 4 & 3.2 & 0.75 \\ 
			\hline
			
			\multirow{5}{*}{Llama 3} 
			& Zero-Shot & 3 & 4 & 3 & 2 & 4 & 3.2 & 0.75 \\ \cline{2-9}
			& Few-Shot & 3 & 4 & 3 & 2 & 4 & 3.2 & 0.75 \\ \cline{2-9}
			& Chain-of-Thought & 4 & 4 & 3 & 2 & \textbf{5} & \textbf{3.6} & 1.02 \\ \cline{2-9}
			& Constraint-based & 3 & 4 & 3 & 2 & 4 & 3.2 & 0.75 \\ \cline{2-9}
			& Role-based & 3 & 4 & 3 & 2 & 4 & 3.2 & 0.75 \\ 
			\hline
			
			\multirow{5}{*}{Mistral} 
			& Zero-Shot & 3 & 3 & 3 & 2 & 4 & 3.0 & 0.63 \\ \cline{2-9}
			& Few-Shot & 4 & 4 & 3 & 2 & \textbf{5} & \textbf{3.6} & 1.02 \\ \cline{2-9}
			& Chain-of-Thought & 3 & 4 & 3 & 2 & 4 & 3.2 & 0.75 \\ \cline{2-9}
			& Constraint-based & 3 & 4 & 3 & 2 & 4 & 3.2 & 0.75 \\ \cline{2-9}
			& Role-based & 3 & 4 & 3 & 2 & 4 & 3.2 & 0.75 \\ 
			\hline
			
			\multirow{5}{*}{Gemma} 
			& Zero-Shot & 2 & 3 & 3 & 3 & 3 & 2.8 & \textbf{0.40} \\ \cline{2-9}
			& Few-Shot & 3 & 3 & 3 & 2 & 4 & 3.0 & 0.63 \\ \cline{2-9}
			& Chain-of-Thought & 3 & 3 & 3 & 2 & 4 & 3.0 & 0.63 \\ \cline{2-9}
			& Constraint-based & 2 & 3 & 3 & 2 & 4 & 2.8 & 0.75 \\ \cline{2-9}
			& Role-based & 3 & 3 & 3 & 2 & 4 & 3.0 & 0.63 \\ 
			\hline
			
			\multirow{5}{*}{Phi-3 Mini} 
			& Zero-Shot & 1 & 4 & 4 & 4 & 2 & 3.0 & 1.22 \\ \cline{2-9}
			& Few-Shot & 2 & 4 & 4 & 4 & 3 & 3.4 & 0.89 \\ \cline{2-9}
			& Chain-of-Thought & 1 & 3 & 4 & 4 & 3 & 3.0 & 1.10 \\ \cline{2-9}
			& Constraint-based & 1 & 3 & 4 & 5 & 2 & 3.0 & 1.22 \\ \cline{2-9}
			& Role-based & 2 & 4 & 3 & 4 & 3 & 3.2 & 0.75 \\ 
			\hline
			\textbf{Average} & -- & 2.6 & 4.0 & 3.0 & 2.3 & 3.8 & 3.1 & 0.8 \\ 
			\hline
		\end{tabular}
	}
\end{table*}

\section{Discussion}
\label{sec:discussion}

Lightweight LLMs show promising potential for analysing user feedback in RE. Although more efficient and transparent than larger models, they are not yet suitable for reliable industrial use without further customization.

\noindent\textbf{A) Feedback Classification.} In feedback classification, Llama 3 and Mistral achieved F1-scores around 0.74 for identifying user request types. This shows that lightweight models can reliably detect explicit requests such as feature suggestions or bug reports. Their performance in NFR classification was weaker, with the best F1-score reaching 0.55. This result aligns with prior studies where models also struggled to capture implicit qualities like usability or reliability~\cite{lu2017automatic}. The limitation likely stems from two factors. Lightweight LLMs find it difficult to interpret subtle, context-dependent expressions. They also lack sufficient exposure to RE-specific terminology and training data. Structured prompting with few-shot or reasoning examples led to only minor improvements. These findings suggest that contextual examples and reasoning cues can enhance understanding. However, overall accuracy remains moderate. In practice, this may produce noisy classifications that mislead analysts. As a result, important feedback can be missed, while irrelevant comments may be misclassified as critical requirements.

\noindent\textbf{B) Requirements Specification Generation}. Lightweight LLMs produced requirement specifications that were generally clear, coherent, and complete. They expressed user feedback readably but often failed to follow formal structures. Their outputs were sometimes verbose and required editing for conciseness and compliance with standards. Although the generated text was fluent, the models occasionally fabricated requirements, adding information absent from the original feedback. As a result, the content appeared plausible but not always accurate or grounded. In practice, these limitations mean lightweight LLMs can assist analysts by drafting initial requirement statements or summarising feedback. However, they still require human review to ensure accuracy and proper structure. With further refinement and adaptation, they could serve as drafting and documentation aids rather than autonomous specification generators.

\noindent\textbf{C) Implications for Requirements Engineering.} Lightweight LLMs can support RE tasks such as feedback filtering, classification, and initial specification drafting. However, they cannot yet replace human analysts. Their moderate precision and recall make them unreliable for use without supervision. These weaknesses may lead to missed insights or false positives that increase review effort.
The generated outputs are clear and coherent but often lack factual grounding and formal consistency. This limits their value for downstream RE tasks, e.g., validation, and traceability. In several tasks, their performance is similar to earlier ML approaches~\cite{Dabrowski2022,lu2017automatic}. Larger models alone do not guarantee better outcomes in RE. Future work should adapt models to RE contexts using prompt design, fine-tuning, and retrieval-based approaches. These methods can improve accuracy and help lightweight LLMs better support early RE tasks.

\section{Threats to Validity}
\label{sec:Threatstovalidity}

\noindent\textbf{Internal Validity.}
The main threat lies in the manual evaluation of generated requirement specifications by a single evaluator, \JD{without validating its reliability}. This introduces potential subjectivity. To mitigate this, we applied a systematic rubric with clear criteria and examples of both high- and low-quality specifications. The rubric was used consistently across all outputs. In addition, prompts and model parameters were standardised to ensure uniform conditions.

\noindent\textbf{External Validity.}
We used two publicly available datasets from different domains to reduce domain bias. Their diverse vocabulary helps generalisability, though an app review represents only one feedback type. Results may not generalise to industrial datasets or other contexts such as issue trackers. Moreover, the study focused on lightweight LLMs; larger models may yield different results.

\noindent\textbf{Construct Validity.}
We employed standard precision, recall, and F1-score metrics for classification and a rubric assessing completeness, consistency, and correctness for generation. As qualitative evaluation relied on one evaluator, some interpretation bias may persist. Prompt phrasing may also influence results; this was mitigated by systematically applying established prompting strategies.

\noindent\textbf{Conclusion Validity.}
All models were tested under identical settings and prompts. However, the limited sample size for specification generation lowers statistical power, and no inferential tests were applied, making results exploratory. While precision and recall were key metrics, tasks may favour one over the other, and F1 may not always be optimal~\cite{Berry2022}. \JD{Our aim was to assess overall practical effectiveness rather than examine differences across RE-specific tasks.}
\section{Related Works}
\label{sec:RelatedWorks}

Online feedback analysis has long supported requirements and software engineering, with many automated methods proposed~\cite{AlSubaihin2021,Martin2017AppSA,Dabrowski2022}. Earlier work relied on traditional ML and NLP methods for tasks such as feedback classification, topic detection, and opinion extraction~\cite{Dabrowski2022,10.1145/3444689}. Our study takes a new direction by applying LLMs to both established classification tasks and a novel one: generating requirements specification~\cite{11190385}, a capability beyond earlier approaches~\cite{Dabrowski2024}. Prior work also highlighted the lack of tools for automated requirements generation from user feedback~\cite{Dabrowski2022a}; we address this gap by evaluating lightweight LLMs across both classification and generation tasks~\cite{cheng2024generative}.

Previous studies rarely ensured reproducibility or revisited results with newer AI methods~\cite{Dabrowski2022}. We emphasise transparent experimentation by benchmarking lightweight models across RE tasks and releasing a full replication package for open, comparative research. While many recent works rely on proprietary models such as GPT~\cite{cheng2024generative,zadenoori2025largelanguagemodelsllms}, we focus on open, lightweight alternatives to improve transparency and reproducibility. Unlike prior LLM-based studies, our work applies lightweight LLMs to online feedback analysis~\cite{zadenoori2025largelanguagemodelsllms}. To our knowledge, this is the first study to evaluate lightweight LLMs for feedback-driven requirements development, supported by a replication package for reproducible RE research.

\section{Conclusion}
\label{sec:conculsion}

Analysing online user feedback is important for RE tasks, yet automation remains challenging. This study presents the first systematic evaluation of open, lightweight LLMs for feedback-driven RE. Five models were evaluated with multiple prompting strategies on three RE tasks: classifying feedback by user request and NFR type, and generating requirement specifications.

Lightweight LLMs classified user request types accurately (F1 $\approx$ 0.74) but performed moderately for NFRs (F1 $\approx$ 0.55). Structured prompting improved results slightly. Generated specifications were generally complete and coherent but often verbose and sometimes included fabricated requirements. Lightweight LLMs can support early RE activities, but still require human oversight.

We release our evaluation framework and replication package to support further research~\cite{mallya2026from}. Future work should explore fine-tuning lightweight LLMs, developing retrieval-augmented pipelines, and creating domain-adapted benchmarks to improve their accuracy and reliability in RE.

\begin{credits}
\subsubsection{\ackname} 

A major part of this work was conducted as part of the MSc thesis of M. A. Mallya, supervised by J. Dąbrowski~\cite{Mallya2025}. The results contribute to the Prompt Me project~\cite{Dabrowski2024}.  This work was funded by Taighde Éireann – Research Ireland (13/RC/2094\_2) and co-funded by the EU under the SyMeCo programme (101081459). The views expressed are those of the authors only.

\subsubsection{\discintname}
The authors have no competing interests to declare that are relevant to the content of this article. \JD{ChatGPT-4~\cite{OpenAI2024ChatGPT4} was used to improve the readability, and the authors remain fully responsible for the final content.}
 
\end{credits}
%
%
%
%

\bibliographystyle{splncs04}
\bibliography{bib_refsq_26}

\end{document}